\documentclass[aps,prl,showpacs,twocolumn,superscriptaddress]{revtex4-1}

\usepackage{graphicx,upgreek,amsmath,amsfonts,textcomp,color}
\usepackage{natbib}

\newcommand{\fref}[1]{Fig.~\ref{#1}}

\newcommand{\BFO}{BiFeO$_3$ }
\newcommand{\PZT}{PZT }
\newcommand{\STO}{SrTiO$_3$ }

\begin{document}
\title{Domain Wall Roughness in Stripe Phase \BFO Thin Films}
\author{B. Ziegler}
\affiliation{DPMC-MaNEP, Universit{\'e} de Gen{\`e}ve, 24 Quai Ernest Ansermet, 1211 Geneva, Switzerland}
\author{K. Martens}
\affiliation{DPMC-MaNEP, Universit{\'e} de Gen{\`e}ve, 24 Quai Ernest Ansermet, 1211 Geneva, Switzerland}
\affiliation{LIPhy, Universit{\'e} Joseph Fourier Grenoble 1, UMR 5588 et CNRS, F-38402 Saint Martin d'H{\`e}res, France}
\author{T. Giamarchi}
\affiliation{DPMC-MaNEP, Universit{\'e} de Gen{\`e}ve, 24 Quai Ernest Ansermet, 1211 Geneva, Switzerland}
\author{P. Paruch}
\affiliation{DPMC-MaNEP, Universit{\'e} de Gen{\`e}ve, 24 Quai Ernest Ansermet, 1211 Geneva, Switzerland}
\date{\today}
\begin{abstract}

Using the model system of ferroelectric domain walls, we explore the effects of long-range dipolar interactions and periodic ordering on the behavior of pinned elastic interfaces. In piezoresponse force microscopy studies of the characteristic roughening of intrinsic 71$^\circ$  stripe domains in \BFO thin films, we find unexpectedly high values of the roughness exponent $\zeta=0.74\pm0.10$, significantly different from those obtained for artificially written domain walls in this and other ferroelectric materials. The large value of the exponent suggests that a random field-dominated pinning, combined with stronger disorder and strain effects due to the step-bunching morphology of the samples, could be the dominant source of pinning in the system.

\end{abstract}
\maketitle

%INTRODUCTION ----------------------------------------------------------------------------------------------------------------------------------------------
The rich static and dynamic physics of pinned elastic interfaces can be understood in terms of  the competition between the flattening effects of elasticity and fluctuations in the potential energy landscape, and describes phenomena as diverse as contact lines \cite{moulinet_pre_04_distribution_width_contact_line}, imbibition fronts \cite{wilkinson_invasion}, vortices in type II superconductors \cite{blatter_94_4mp_vortex_review}, fracture propagation \cite{maloy_prl_92_crack}, magnetic domain walls \cite{lemerle_prl_98_FMDW_creep}, and surface growth \cite{Barabasi_surface_growth}.  Ferroelectric domain walls provide a useful model system in which many aspects of such glassy behavior can be readily accessed \cite{giamarchi_domainwall_review}. Previous studies of roughening, nonlinear dynamics, and aging have focused primarily on individual domain walls in uniaxial materials \cite{paruch_cras_13_DW_review}. However, a particularly interesting experimental and theoretical challenge is posed by systems where coupled ferroic orders (such as ferroelectricity and ferroelasticity, or ferroelectricity and (anti)ferromagnetism \cite{daraktchiev_prb_10_BFO_DW}), as well as long-range interactions could lead to more-complex behavior. 

Room-temperature multiferroic \BFO is an excellent candidate for investigating such phenomena. In this perovskite, polarization orientation along the eight pseudocubic [111] axes gives rise to three domain wall types (180$^\circ$, purely ferroelectric, and 71$^\circ$, 109$^\circ$, also ferroelastic), with magnetoelectric coupling between the ferroelectric and antiferromagnetic orders \cite{chu_natmat_10_BFO_ME}. In addition, unusual domain wall functionalities  \cite{seidel_natmat_09_BFO,martin_nl_08_xchange_BFO} hold promise for future nanoelectronic applications \cite{bea_natmat_09_domainwalls_NV,catalan_rmp_12_DW_review}. \BFO films with specific polarization orientations, and domain structures ranging from nanoscale, sometimes fractal-like ``bubbles'' to well-defined stripes can be obtained by adapting the deposition conditions and substrate \cite{Chu:2009dg,Das:2006by,daumont_PRB_10_BFO}. Artificial domains can also be written by a biased scanning probe microscopy (SPM) tip, although this procedure can introduce significant electrochemical changes \cite{kalinin_nano_11_electrochemical_SPM}. Intrinsic stripe domains follow standard Landau-Lifshitz-Kittel scaling of domain period $w \sim h^{1/2}$ with the sample thickness $h$ \cite{chen_apl_07_BFO_domain_width}, while in samples with fractal bubble domains, a modified $0.59$ exponent and apparent one-dimensional roughening of artificial domains were observed \cite{catalan_prl_08_BFO_DW}. 

Previous studies considered the domain walls as individual interfaces weakly pinned by disorder, with monoaffine roughness scaling characterized by a single-valued roughness exponent $\zeta$, dependent only on the disorder universality class and the system dimensionality. However, the heterogeneous disorder of ferroelectric thin films, with local universality class fluctuations and strong pinning \cite{jesse_natmat_08_SSPFM}, has recently been shown to lead to a breakdown of monoaffinity \cite{guyonnet_prl_12_multiscaling}. Moreover, in periodic systems interactions between neighboring interfaces can limit roughening and change the effective dimensionality \cite{giamarchi_vortex_review}. To better understand the roughening of \BFO domain walls, the possible effects of interactions, complex domain structure, and a heterogeneous, potentially dynamic disorder landscape must therefore be considered.

In this Letter, we report on a direct comparison of the roughening of individual, well-separated artificial domain walls written in bubble domain films with that of intrinsic, periodic stripe domains. We find a roughness exponent $\zeta=0.48\pm0.09$ for the former, in good agreement with previous measurements \cite{catalan_prl_08_BFO_DW}. Intrinsic stripe domains show much higher $\zeta=0.74\pm0.10$ values, possibly reflecting the dominant effects of random field pinning.

% MATERIALS AND METHODS ---------------------------------------------------------------------------------------------------------------------------
\begin{figure*}
\includegraphics[width=\linewidth]{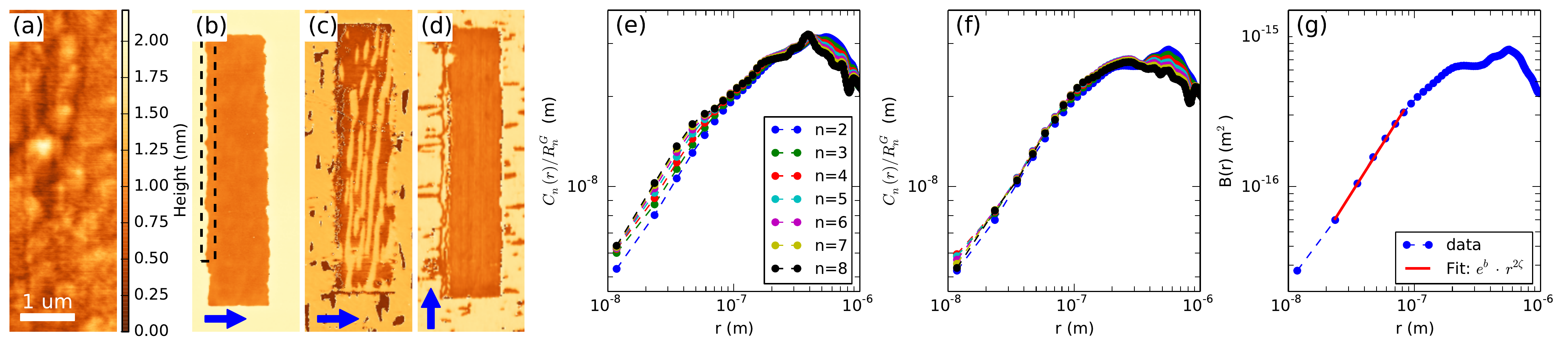}
\caption{Topography (a), vertical (b), and lateral (c,d) PFM phase images of an artificial rectangular domain in the as-grown type I film. Blue arrows indicate cantilever orientation. (e) Renormalized n=2-8 displacement-displacement correlation functions for the full left domain wall show fanning. (f) The correlation functions collapse when only the boxed segment is considered. (g) Roughness function of the boxed segment, with roughness exponent $\zeta$ extracted by fitting (solid (red) line).}
\label{fig:type1}
\end{figure*}

\begin{figure}
\includegraphics[width=\linewidth]{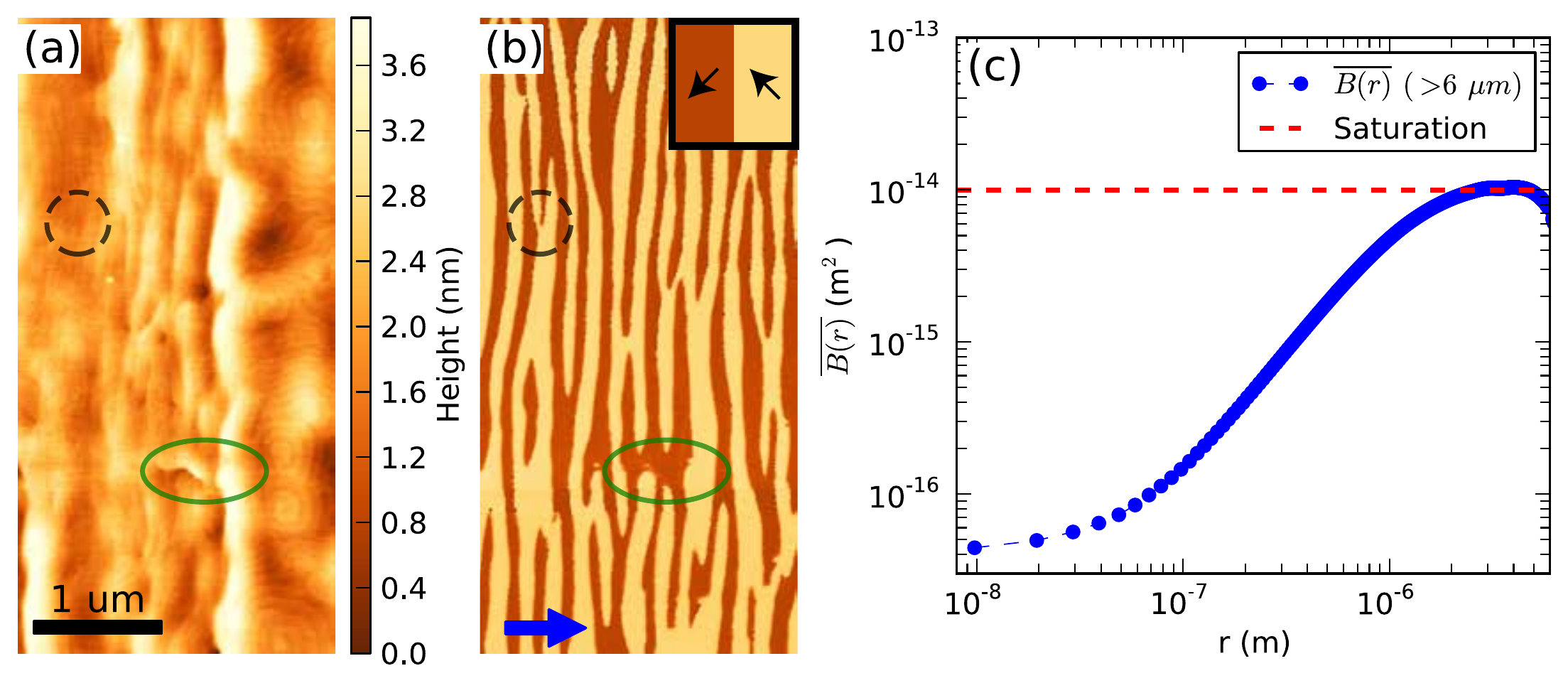}
\caption{Simultaneously obtained (a) topography and (b) lateral phase images of step bunching and intrinsic 71$^\circ$ stripe domains. Blue arrow indicates cantilever orientation. The inset in (b) indicates the in-plane polarization orientation. (c) Average roughness $\overline{B(r)}$ for domain walls $>6\,\upmu$m, with saturation of relative displacements at $\sim$100$\,$nm (dashed red line).}
\label{fig:type2}
\end{figure}

\BFO thin films were epitaxially grown by radio frequency off-axis magnetron sputtering on 10$\,$nm LaNiO$_3$ electrodes on TiO$_2$ terminated (001) \STO single-crystal substrates. Adapting the deposition conditions, thinner type I films with bubble domains, and thicker type II films with stripe domains were obtained \footnote{LaNiO$_3$ electrodes were grown at 550$^\circ$C, 100$\,$mTorr (Ar/O$_2$ = 70/30) immediately prior to \BFO deposition. Type I films were grown at 600$^\circ$C, 150$\,$mTorr (Ar/O$_2 =$95/5), 60$\,$W/cm$^2$ gun power. Type II films were grown at 650$^\circ$C, 100$\,$mTorr (Ar/O$_2$ = 97/3), 100$\,$W/cm$^2$ gun power. SPM measurements were carried out on {\it Asylum Research Cypher}, {\it Veeco Multimode} with {\it Nanonis} controller, with different commercial metallized tips, at 0.1$\,$kHz$-1.1\,$MHz.}.
The films used present 0.2 and 0.5$\,$nm rms surface roughness, respectively, with unit cell steps and some granularity in the 60$\,$nm type I film (\fref{fig:type1}(a)), and a step-bunching morphology in the 300$\,$nm type II film (\fref{fig:type2}(a)) as previously seen in thick \BFO \cite{Das:2006by}. Piezoresponse force microscopy (PFM) reveals four in-plane polarization variants in type I films, and thus 71$^\circ$ and 109$^\circ$ domain walls, with small, somewhat irregular and elongated bubble domains, as can be seen in \fref{fig:type1}(b)-\fref{fig:type1}(d). Type II films present a stripe pattern with only two in-plane polarization variants, and thus only 71$^\circ$ domain walls, as shown in \fref{fig:type2}(b).

To quantify the roughness scaling for intrinsic and artificial domain structures, we carried out detailed PFM measurements in both samples. In the type I film, we wrote artificial rectangular domains with negative SPM tip bias, switching the out-of-plane polarization component to obtain clearly defined noninteracting domain walls (\fref{fig:type1}(b)) separated by at least 1$\,\upmu$m. The in-plane polarization is also modified \cite{balke_natnano_09_BFO,bea_jpcm_11_BFO_switching} (\fref{fig:type1}(c) and \fref{fig:type1}(d)), giving a complex, mixed 71$^\circ$/109$^\circ$/180$^\circ$ character across the domain wall.  SPM writing also affects the sample surface, inducing initially strong vertical PFM amplitude contrast, which gradually decays with time, and permanent raised features or particulates in the written areas. We attribute these effects to a combination of reversible surface charging and reorientation of mobile defects, and minor irreversible surface damage during switching. In contrast to these artificial structures, the intrinsic 71$^\circ$ walls of long (up to 20$\,\upmu$m) stripe domains in the type II film, formed during sample growth without exposure to high intensity switching fields, show no such surface modification. The stripe domains present a clear periodicity $w \sim$100$\,$nm and preferred orientation, globally aligned with the step-bunching direction, as well as obvious wandering and tapering off or branching of individual domains.

To analyze the roughening of artificial or intrinsic domain walls, we first map their position from binarized vertical or lateral PFM phase images, respectively. For a given length scale $r$, we extract the relative displacements from the elastically optimal reference configuration, determined by least squares fitting, as $\Delta u(r)=u(z)-u(z+r)$, where $u(z)$ are the transverse displacements at longitudinal coordinate $z$ along the wall. The characteristic roughness scaling can be obtained from the central moments
\begin{equation}
M_n(r) = \overline{\langle | \Delta u(r)|^n \rangle} \propto r^{n\,\zeta_n}
\end{equation}
of the probability distribution function of these relative displacements or from the related correlation functions 
\begin{equation}
C_n(r) = \overline{\langle | \Delta u(r) |^n \rangle^{1/n}} \propto r^{\zeta_n}
\end{equation}
where $\langle\,\cdots\rangle$ and  $\overline{\cdots}$ average over $z$, and over different disorder realizations, respectively, and $\zeta_n$ is the $n$th moment scaling exponent \cite{agoritsas_physb_12_DES}. For weak disorder pinning, interfaces are well described by a monoaffine Gaussian probability distribution function \cite{rosso_jstatmech_05_gaussian},  and the $C_n(r)$ collapse to a universal curve when normalized by the $r$- and $\zeta$-independent Gaussian ratio $R_n^G = C_n^G(r)/C_2^G(r)$ \cite{santucci_pre_07_fracture_statistics}. We applied this test to n=2-8 orders.

% RESULTS AND DISCUSSION ---------------------------------------------------------------------------------------------------------------------------
For the type I film, we find that for $\sim$40\% of the artificial domain walls monoaffine scaling breaks down when their full length (5-6$\,\upmu$m) is considered, as shown in \fref{fig:type1}(e), with clear fanning and offset of the Gaussian-normalized $C_n(r)$. Since the roughness function $B(r) = M_2(r) \propto r^{2\,\zeta}$ can only be defined for monoaffine scaling, this observation underlines the importance of such analysis \emph{before} extracting a universal roughness exponent $\zeta_n = \zeta \,\forall\, n$.  As for artificial Pb(Zr$_{0.2}$Ti$_{0.8}$)O$_3$ (\PZT) domain walls \cite{guyonnet_prl_12_multiscaling}, the breakdown of monoaffinity can be related to highly localized features leading to strong fluctuations of the domain wall position. Their exclusion allows monoaffinity to be recovered (\fref{fig:type1}(f)). Considering 43 monoaffine domain wall segments 0.8-5.9$\,\upmu$m long for sufficient statistics, we then determined $B(r)$ (\fref{fig:type1}(g)). As a result of the artificial writing process with linear SPM tip motion \cite{paruch_dw_review_07}, $B(r)$ rapidly saturates at $r^*\sim100\,$nm, with equilibration to power-law roughness scaling only at short length scales, from which $\zeta$ can be extracted. Averaging $B(r)$ yields $\zeta_{avg}=0.51$,  while averaging individual roughness exponents yields $\bar{\zeta} = 0.48 \pm 0.09$. These values are close to measurements of artificial domain walls in \BFO ($\zeta_{avg}=0.56$)\cite{catalan_prl_08_BFO_DW} and \PZT ($\bar{\zeta}=0.57\pm0.05$)\cite{guyonnet_prl_12_multiscaling}.

\begin{figure}
\includegraphics[width=\linewidth]{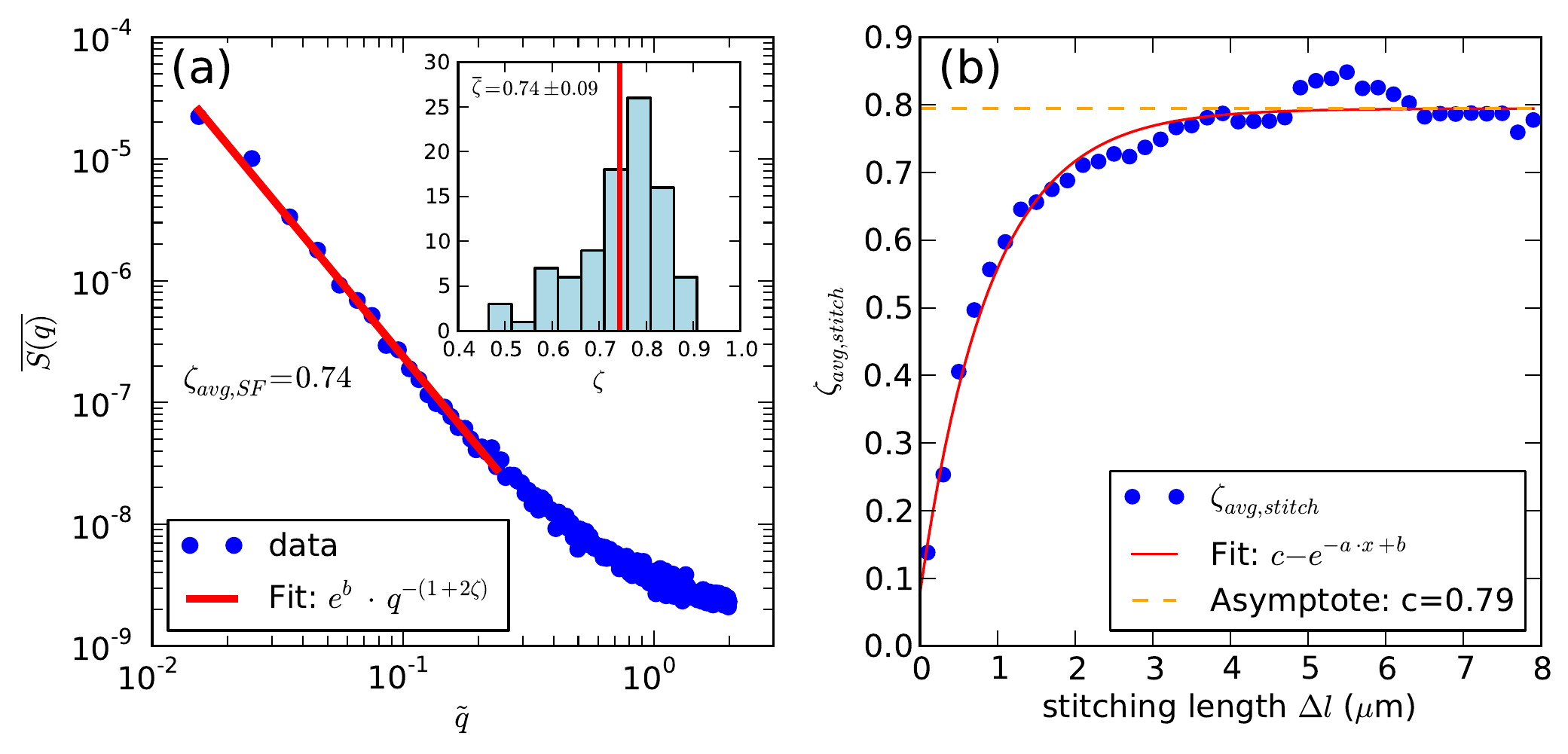}
\caption{(a) Average structure factor $S(q)$ versus the corrected wave vector $\tilde{q}$ for 92 monoaffine intrinsic 71$^\circ$ domain walls, with roughness exponent fitting (solid red line). Inset: Histogram of the corresponding individual roughness exponents $\zeta$, with mean $\bar\zeta$ indicated by the (solid red) line. (b) Roughness exponent $\zeta$ as a function of stitching length, with saturation value (dashed orange line) determined by fitting (solid red line).}
\label{fig:histo}
\end{figure}

The intrinsic 71$^\circ$ domain walls in the type II film present a rather different behavior. Although once again we observe localized breakdown of monoaffinity, this can be related to structural features (\fref{fig:type2}(a) and \fref{fig:type2}(b) solid (green) ellipse) or to branching or avoidance interactions between different stripe domains (\fref{fig:type2}(a) and \fref{fig:type2}(b) dashed (black) circle). Considering $\overline{B(r)}$ (\fref{fig:type2}(c)) for the longest (6-16$\,\upmu$m) monoaffine domain wall segments, from which such features have been excluded, we find an extensive power-law growth region, with an unexpectedly high $\zeta_{avg} = 0.74$. Averaging individual roughness exponents likewise yields $\bar\zeta=0.74 \pm 0.10$, as shown in the inset of \fref{fig:histo}(a) for all 92 monoaffine domain wall segments, ranging from 2.5 to 15.9$\,\upmu$m. We also observe a flattening at the lowest length scales, and saturation for $r > 2.5\,\upmu$m. While we believe the flattening at lowest length scales to be an artifact of the finite imaging resolution \cite{Schmittbuhl:1995wr} \footnote{In both repeated measurements of the same wall with lower resolution, and digital ``coarse-graining'' of a high resolution image after measurement, the flattening appears at higher length scales and progressively higher $B(r)$ values as a function of pixel size, effectively decreasing $\zeta$ and rounding the crossover between the flattening and power-law-growth regions (see Supplemental Material). To determine $\zeta$, the highest resolution ($\sim$10$\,$nm per pixel) was therefore used.}, the saturation at $r > 2.5\,\upmu$m appears robust, and cannot be a writing artifact, since the intrinsic domains are formed during high-temperature growth, allowing better equilibration with the disorder landscape.  

To better understand the roughening behavior, we need to consider the complex experimental situation in BiFeO$_3$ thin films, where combined effects of long-range interactions, both dipolar and strain mediated, as well as the nature of the disorder and the periodicity of the system  for the case of the 71$^\circ$ stripe domains need to be explicitly addressed.

For randomly pinned \emph{periodic} systems, above a critical length scale $r_P$  interactions limit the relative displacements of the individual interfaces or manifolds to below the system periodicity, leading to a much slower growth of $B(r) \sim \log(r)$ \cite{giamarchi_vortex_review}. Thus, the observed saturation of $B(r)$ corresponding to relative displacements of $\sim$100$\,$nm, comparable to the stripe domain periodicity, could reflect a crossover to a regime dominated by interactions between neighboring domain walls.  However, in this case, the roughness exponent measured in the random manifold regime at small length scales below $r_P$, where individual interfaces freely wander in the disorder landscape, would reflect the higher effective dimensionality of the system, incompatible with the high exponent value observed for the 71$^\circ$ domain walls.

Modification of the expected roughness exponent value is also possible when long-range interactions ``stiffen'' the interface along one or more directions. For ferroelectric domain walls, dipolar interactions were shown to increase the effective dimensionality, leading to lower values of the roughness exponents \cite{paruch_prl_05_dw_roughness_FE,Xiao:2013dy}.
Considering dipolar forces as a first approach towards long-range effects, we find that the effective elastic energy term contains two parts: one related to the Ginzburg gradient term, quadratic in the local variation of polarization across the domain wall \cite{catalan_rmp_12_DW_review}  leading to a standard short-range $q^2$ elasticity, the second related to the dipolar interactions with a more complex form (see Supplemental Material). The angle which minimizes the dipolar energy corresponds exactly to that between the neighboring in-plane pseudocubic orientations, as observed experimentally via lateral PFM (\fref{fig:type2}(b)). The analysis of the dipolar energy for $71^\circ$ domain walls in \BFO shows that the resulting elastic energy of the wall is anisotropic in its dependence on the wave vector of the domain wall fluctuations with respect to the out-of-plane direction (see Supplemental Material). 
Interestingly, the resulting effective dimensionality depends on the angle between the in-plane polarization components across the domain wall. For isolated domain walls, dominant dipolar interactions would lead to a $|q|$ elastic coefficient along the wall, and thus to roughness $\zeta_{RB,d=1,D} = \zeta_{RB,d=2,D} = 0.2$, for both one-dimensional and two-dimensional walls, as compared to $\zeta_{RB,d=1,E} = 2/3$ and $\zeta_{RB,d=2,E} \simeq 0.4$ for the standard $q^2$ elasticity. For random-field disorder, the exponent for a dipolar-interaction-dominated elasticity would be $\zeta_{RF,d=1,D} = \zeta_{RF,d=2,D} = 1/3$ and $\zeta_{RF,d=1,E} = 1$, $\zeta_{RF,d=2,E} = 2/3$ for a standard $q^2$ elasticity. Although none of these values directly correspond to the experimental observations, the closer match appears to be with random-field disorder in the one-dimensional case, since the random-bond exponent values are systematically lower than those observed experimentally. Random fields with a crossover between a dipolar-dominated elasticity and a standard (short-range) elasticity could thus lead to an exponent compatible with the experimental value. Of course other effects must be considered as well, in particular the effects of long-range strain interactions, so the full theoretical interpretation of the observed exponent is extremely challenging. In any case, given the wide range of possible roughness exponents, above all it is crucial to determine the value of $\zeta$ as accurately as possible.

This value can be strongly under- or overestimated by different analysis methods \cite{Schmittbuhl:1995wr}, and is especially problematic for values close to $\zeta\in\mathbb{N}$. To verify the high $\zeta$ value we obtained from the roughness function $B(r)$, we therefore used the Fourier transform $\hat{u}(q)$ of the displacement $u(r) - \bar{u}$ from its mean value $\bar{u}$ to compute the monoaffine structure factor \cite{Schmittbuhl:1995wp,Rosso:2007dj, Kolton:2009iz}:
\begin{equation}
S(q) = |\hat{u}(q)|^2 \propto q^{-(1+2\zeta)}
\end{equation}
This method \footnote{For the amplitude $\hat{u}_k$ of the Fourier mode $q_k$, we take the convention $\hat{u}_k = \sum\limits_{j=0}^{N-1} (u_j-\bar{u})e^{-q_k\,j}$, where $\bar{u}$ is the mean value, $q_k = \frac{2\pi k}{N}$, and $N$ the total number of pixels. For a discrete Fourier transformation the structure factor $S(q)$ is plotted against a corrected $q$ vector, $\tilde{q}=2\sin\left(\frac{\pi k}{N}\right)$, for linearisation in a log-log scale (as used in \cite{Rosso:2007dj,Kolton:2009iz}).} not only returns more reliable $\zeta$ values, but also allows $\zeta\in\mathbb{R}$, which is especially important if $\zeta$$\rightarrow$1. In contrast, the roughness function $B(r)$ limits $\zeta$ values between 0 and 1. Extracted from the average structure factor $\overline{S(q)}$ for the same set of 92 walls as in the inset in \fref{fig:histo}(a), the roughness exponent $\zeta_{avg,SF} = 0.74$ nonetheless confirms the high value. We attribute the flattening at higher $\tilde{q}$ values to the same tip resolution limitation which led to a saturation of $B(r)$ at lower length scales $r$ \cite{Schmittbuhl:1995wr}.

\begin{figure}
\includegraphics[width=\linewidth]{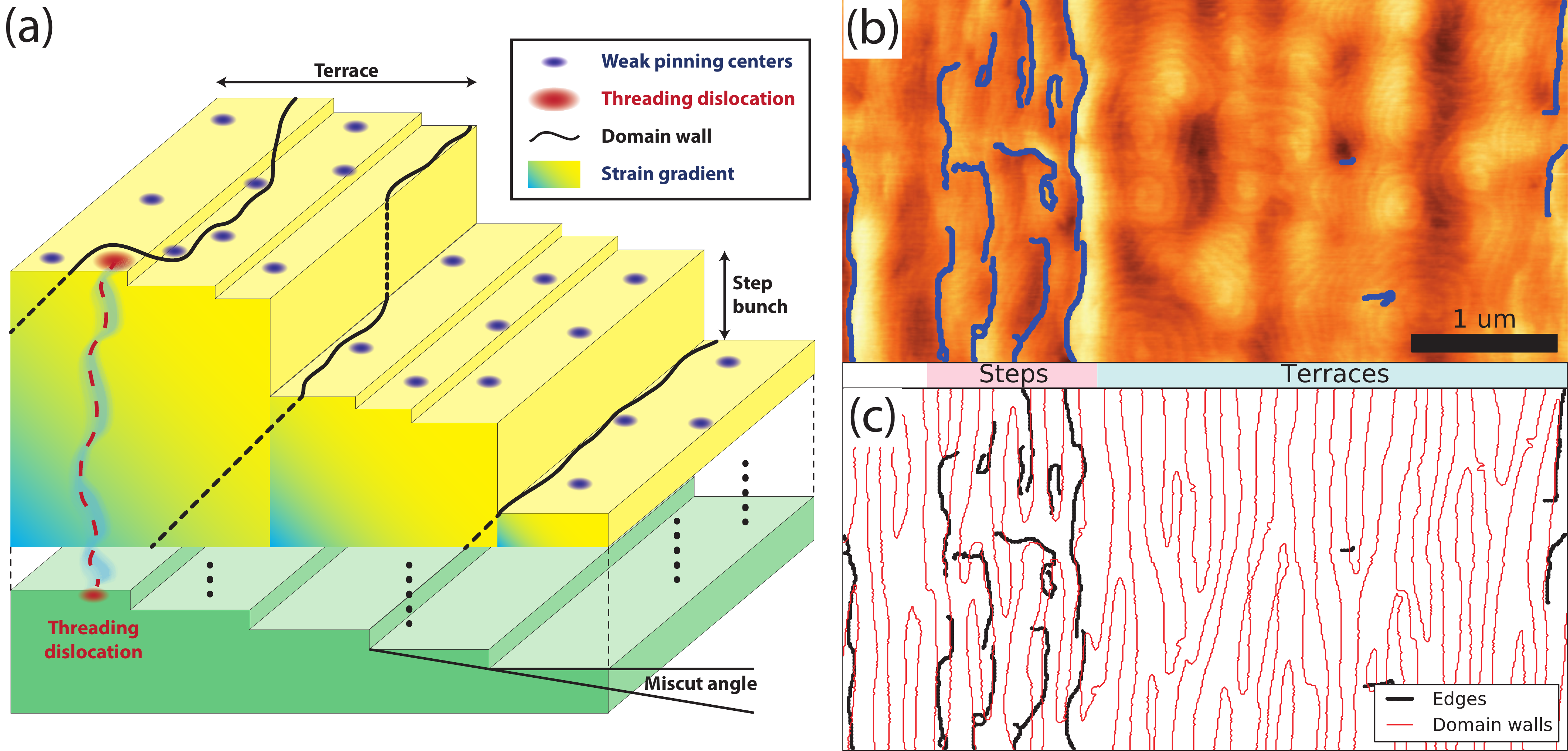}
\caption{(a) Schematic diagram of different possible pinning centers in a \BFO film with step bunches (larger steps) and terraces with unit cell steps (smaller steps). 71$^\circ$ domain walls are drawn as solid (black) lines on top of the film and dashed (black) lines into the film. The small (dark blue) spots and the larger (dark red) spots indicate weak pinning centers and threading dislocations, respectively. The color gradient indicates the strain gradient introduced by the substrate's vicinality, whereas the blue (dark) and the yellow (bright) areas indicate the strained and relaxed \BFO regions. (b) Topography image of the type II BiFeO$_3$ film overlaid with the edges of the step bunches (dark lines). Areas of higher step density and terraces are marked by the red and blue bar below the figure. (c) Overlay of the step-bunch edges (dark lines) with the domain walls (red lines) extracted from the lateral phase image of the same region as (b).}
\label{fig:overlay}
\end{figure}
One additional important feature of the type II sample is the step bunching morphology and its potential interaction with domain walls. Step bunching on vicinal \STO substrates was shown to generate anisotropic strain gradients, introducing nonuniform local internal fields \cite{kim_apl_09_BFO_DW_step_bunching}, and was also associated with vertical misfit dislocations \cite{kim_apl_11_BFO_step_bunch}. Such features could give rise to strong correlated pinning, in particular of ferroelastic ferroelectric domain walls, for which the simple assumption of random disorder may no longer be valid, and more appropriate theoretical models including the complex internal structure and energy landscape of \BFO, schematically represented in \fref{fig:overlay}(a), may need to be considered.
However, as can be seen in \fref{fig:overlay}(c), we observe that while the stripe domains (blue lines) generally align with the step bunching direction (step edges marked in black, from \fref{fig:overlay}(b)), and domain walls follow particularly steep step edges for 1-3 $\upmu$m segments, they also in many cases traverse across them, suggesting that some degree of randomness is still present. Regions with a high density of step bunching, as well as wide terraces with unit-cell steps show similar distribution and roughness of domain walls.

To further explore the possibility of correlation between the step bunching morphology and domain wall roughening, we used a simple windowing method to exclude larger excursions in $u(r)$, possibly arising from such a correlation, and from branching/avoidance between domains. Splitting every individual domain wall into segments of length $\Delta l$, we subtracted the slope with respect to the first and last point for each segment, then artificially restitched them. This procedure allows roughening to be analyzed as a function of $\Delta l$, beyond which correlations in the relative displacements are effectively removed, while retaining high statistics for sufficient averaging.  Distinct regimes such as short length scale roughening due to disorder crossing over to strong correlated pinning at large length scales should lead to different roughness exponents. We would, therefore, expect distinct plateaus in the evolution of $B_{\Delta l}(r)$ and $\zeta_{\rm{stitch}}$. However, \fref{fig:histo}(b) shows rapidly increasing $\zeta_{\rm{avg,stitch}}$ with only one saturation regime for $\Delta l>2.5\,\upmu$m (blue circles), and a saturation value of $\zeta_{\rm{sat}}=0.79$ as an upper bound (dashed orange line), indicating no second length scale regime with a different roughness value.

%Conclusion
Taken together, these results suggest that the 71$^\circ$ domain walls in the striped \BFO phase act as individual random manifolds over the length scales of observation, with a single roughening regime characterized by a high-value roughness exponent $\zeta = 0.74\pm0.10$.

One scenario, compatible with the observed value, would be that of random-field pinning of one-dimensional domain walls.
In the \BFO samples, the quasiperiodic step-bunching morphology which could potentially provide strong correlated pinning lines and influences the global alignment of the domain walls, appears nonetheless to allow some randomness in their actual position, but could perhaps contribute to a field-dominated pinning.  However, to ascertain the actual microscopic mechanism, further studies taking into account the complexity of \BFO and its domain walls both theoretically and experimentally are needed. We see our results as a strong motivation for such further investigation of the roughness behavior of ferroelectric domain walls in the presence of strong and/or correlated disorder/defects.

%ACKNOWLEDGEMENTS
\begin{acknowledgments}
The authors thank S. Santucci for helpful discussion, M. Lopes and S. Muller for technical support, the European Commission FP7 OxIDes, and Swiss National Science Foundation for financial support under the NCCR MaNEP and Division II. T. G. acknowledges NSF Grant No. 1066293 and thanks the Aspen Center for Physics for hospitality during the writing of this Letter. K.M. acknowledges the excellence scholarship of the University of Geneva.\\
Corresponding author B. Z.~(benedikt.ziegler@unige.ch)
\end{acknowledgments}

%\bibliography{PPtot}
%\bibliography{BZtot}

%

\end{document}